# Excellent Thermoelectric and Piezoelectric Properties of Differently Stacked Layers of Two-Dimensional Transition Metal Dinitride HfN$_2$

Atanu Betal, Jayanta Bera, Satyajit Sahu*

*Department of Physics, Indian Institute of Technology Jodhpur, Jodhpur 342037, India*

## *Abstract*

Two-dimensional (2D) transition metal dinitride (HfN$_2$) has been studied for their optoelectronic, piezoelectric, and thermoelectric properties. Both monolayer and bilayer of HfN$_2$ were studied using density functional theory (DFT) and Boltzmann transport equation (BTE). Bilayer of HfN$_2$ with different stacking layers (AA and AB) showed different electronic properties. The optical property of the material suggests that it is a very good absorber in the ultraviolet (UV) region thus, can be used as a UV-photodetector and as an absorber layer in photovoltaic devices. The piezoelectric properties of the material also showed promising behavior as the piezoelectric stress and strain tensors have highest value of $8.97 \times 10^{-10} C/m$ and 12.59 pm/V respectively for the bilayer. The piezoelectric tensors have highest value for AB stacked bilayer. The ZT value of 0.8 at 900 K is also highest for bilayer AB stacked HfN$_2$. These high values of piezoelectric and thermoelectric parameters of the material suggest that the material would be an excellent choice as thermoelectric energy harvesting devices as well as mechanical stress sensor or actuator.

# Introduction:

The demand of clean energy is increasing because of exhaustion of non-renewable energy sources, for that piezoelectric and thermoelectric devices can be thought of trusted options. Piezoelectric materials are those, which can convert mechanical energy to electrical energy or vice-versa. These are useful in mechanical stress sensors, actuators[1,2]. Thermoelectric material converts thermal energy into electrical energy which depends on the seebeck coefficient (S), electrical conductivity (σ), temperature (T) and thermal conductivity (K) of the material. The advent of graphene, the "hottest star" of two dimension (2D) materials has ignited great interest and opened wide research areas to study the properties of 2D materials[3–5]. However, the semi metallic property of graphene obstructed its application in microelectronic devices[6]. In search of 2D material with wide band gap people found many layered materials including hexagonal boron nitride (h-BN)[7–9], transition metal oxide[10,11], transition metal dichalcogenides (TMDCs)[12], silicene[13], germenene[14] etc. These 2D materials are promising candidates for piezoelectric and thermoelectric devices. Among all TMDCs monolayer $MoS_2$, $MoSe_2$ are the most popular materials with direct band gap and have potential to be used as energy storage, thermoelectric, piezoelectric, gas sensor, energy harvesting and many more electronic devices[15–19]. Another TMDC $WS_2$ also shows good thermoelectric behavior at high and moderate temperatures with applied strain[20,21]. Apart from TMDC, metal diiodides also show extraordinary thermoelectric properties with ultra-low lattice thermal conductivity[22]. Black phosphorus monolayer called as phosphorene also attracted attention because of its extraordinary electronic properties as it has a moderate direct bandgap which is useful for optoelectronic devices, although it has stability issue in air[23,24].

Recently, transition metal dinitride materials (TMDNs)are being studied because of their astonishing physical properties. . After successful synthesis of $TiN_2$ having tetragonal structure as predicted previously by DFT, at high pressure using TiN and dense $N_2$ gas as precursors we were motivated to study about TMDNs[25,26]. Unlike graphene, half metallic $YN_2$ monolayer shows ferromagnetic properties and is a potential candidate for spintronic devices[27]. Two dimensional $MoN_2$ were synthesized via solid state ion exchange route and it shows very good catalytic activity and high hydrogenation selectivity which is more than $MoS_2$[28]. It also shows metallic ferromagnetic behavior above room temperature with a Curie temperature at 420K, the highest measured Curie temperature for 2D materials[29]. Bulk $HfN_2$ shows very good ductility and "Through silicon via" technology (TSV) thin film of $HfN_x$ were prepared successfully[30]. Although people are trying to synthesize 2D $HfN_2$ the stability of $HfN_2$ monolayer was already predicted theoretically[31].

Here we extend the theoretical studies of $HfN_2$ monolayer as well as bilayer with different stacking configuration. Using different stacking for the bilayer $HfN_2$ the electronic, optical, piezoelectric and thermoelectric properties of the materials were studied theoretically. Both the stacking showed direct band gap at high symmetric K point. High absorption coefficient of $HfN_2$ in the ultraviolet energy region predicts the possible use of the material as UV photodetectors. Piezoelectric properties of the materials were also studied and they showed very good piezoelectric effect with high piezoelectric tensors. Highest piezoelectric stress and strain tensor were shown by $HfN_2$ bilayer with AB stacking, and the values are $8.97 \times 10^{-10}$ C/m and 12.59 pm/V. Thermoelectric properties are also calculated for the material. The thermoelectric figure of merit is highest for AB stacked bilayer with highest value for n-type carrier is 0.80 at 900 K. Highest ZT value and high piezoelectric tensor suggest that the material can be used in uv-photodetector thermoelectric device, energy harvesting devices as well as a stress sensor.

## *Calculation details:*

First principle calculations of monolayer and bilayer $HfN_2$ with AA and AB atomic stacking were performed using DFT. Structural optimization, electronic properties were calculated using ultrasoft pseudopotential[32] with nonlinear core correction and Perdew-Burke-Ernzerhof (PBE) functional[33] as implemented in QUANTUM ESPRESSO (QE) package[34]. Van der Waals corrections of Grimme (DFT-D2) were taken into account while calculating the properties of bilayer. The calculations were performed with 15×15×2 and 15×15×4 k-mesh sampling for monolayer and bilayer respectively with Methfessel-Paxton first order spreading. Atoms were relaxed until force convergence minimum of $10^{-3}$ Ry/bohr has been attained. 50 Ry cut-off energy ofplane wave was taken to perform the calculations. SIESTA package[35] was used to calculate optical properties. The momentum space formulation and Kramers-Kronig transformation [36] were used to calculate imaginary ($\varepsilon_2$) and real part ($\varepsilon_1$) of dielectric function respectively. Electronic polarization of mono and bilayer were calculated using geometric Barry phase approach[37] as implemented in SIESTA. The electronic polarization can be given by

$$P_{e,\parallel} = \frac{ife}{8\pi^3} \int dk_\perp \sum_{n=1}^{M} \int_0^{G_\parallel} dk_\parallel \left\langle u_{kn} \left| \frac{\delta}{\delta k_\parallel} \right| u_{kn} \right\rangle \tag{1}$$

e, f, M, $G_\parallel$, $u_{kn}$ are the electron charge, the occupation having value of 2 for non-magnetic system, occupied band number, reciprocal vector along the specified direction and Bloch function, respectively. BoltzTraP code was used to calculate thermoelectric properties of materials[38]. Thermoelectric parameters can be obtained from the equations as stated below

$$\sigma_{\alpha,\beta} = \frac{1}{\Omega} \int \sigma_{\alpha,\beta}(\varepsilon) \left[-\frac{\partial f_\mu(T,\varepsilon)}{\partial \varepsilon}\right] d\varepsilon \tag{2}$$

$$k_{\alpha,\beta}(T,\mu) = \frac{1}{e^2 T \Omega} \int \sigma_{\alpha,\beta}(\varepsilon)(\varepsilon - \mu)^2 \left[-\frac{\partial f_\mu(T,\varepsilon)}{\partial \varepsilon}\right] d\varepsilon \tag{3}$$

$$S_{\alpha,\beta}(T,\mu) = \frac{(\sigma^{-1})_{\gamma,\alpha}}{eT\Omega} \int \sigma_{\gamma,\beta}(\varepsilon)(\varepsilon - \mu) \left[-\frac{\partial f_\mu(T,\varepsilon)}{\partial \varepsilon}\right] d\varepsilon \tag{4}$$

Where, $\sigma_{\alpha,\beta}$, $k_{\alpha,\beta}$, $S_{\alpha,\beta}$, $\Omega$ and $\mu$ are the electrical conductivity, electronic thermal conductivity, Seebeck coefficient, volume of the unit cell and chemical potential respectively. Phonon thermal conductivity, mean free path, phonon relaxation time were calculated using Phono3py[39]. Supercell of 2×2×1 was taken and self-consistent calculations were carried out with 8×8×3 k-mesh and 0.06Å displacement.

## Result and Discussions

### *Structural details:*

HfN$_2$ has hexagonal geometry with P$\bar{6}$m2 symmetry and space group number 187 in which trigonal prismatic structure were formed using one Hf atom and six N atoms with Hf atom as the centre atom. In AA stacking one Hf atom stacks on another Hf atom and similar stacking for N atoms as well whereas in AB stacking N atoms stack on Hf atoms as shown in figure 1(a) and 1(b) respectively. The side and top view of HfN$_2$ bilayer with AA and AB stacking is shown in figure 1. After optimization we get the lattice parameters for unit cell of monolayer and bilayer to be a=b=3.42Å which matched well with previously reported paper[31]. A lattice constant of 25Å along the z direction was considered to sufficiently isolate the layers. The bond lengths of Hf-N and N-N are 2.0864Å and 1.4132Å for monolayer whereas values of Hf-N, N-N and Hf-Hf

distances are 2.1362Å, 1.6564Å, 4.178Å and 2.1390Å, 1.6577Å, 4.9493Å for AA and AB bilayer respectively.

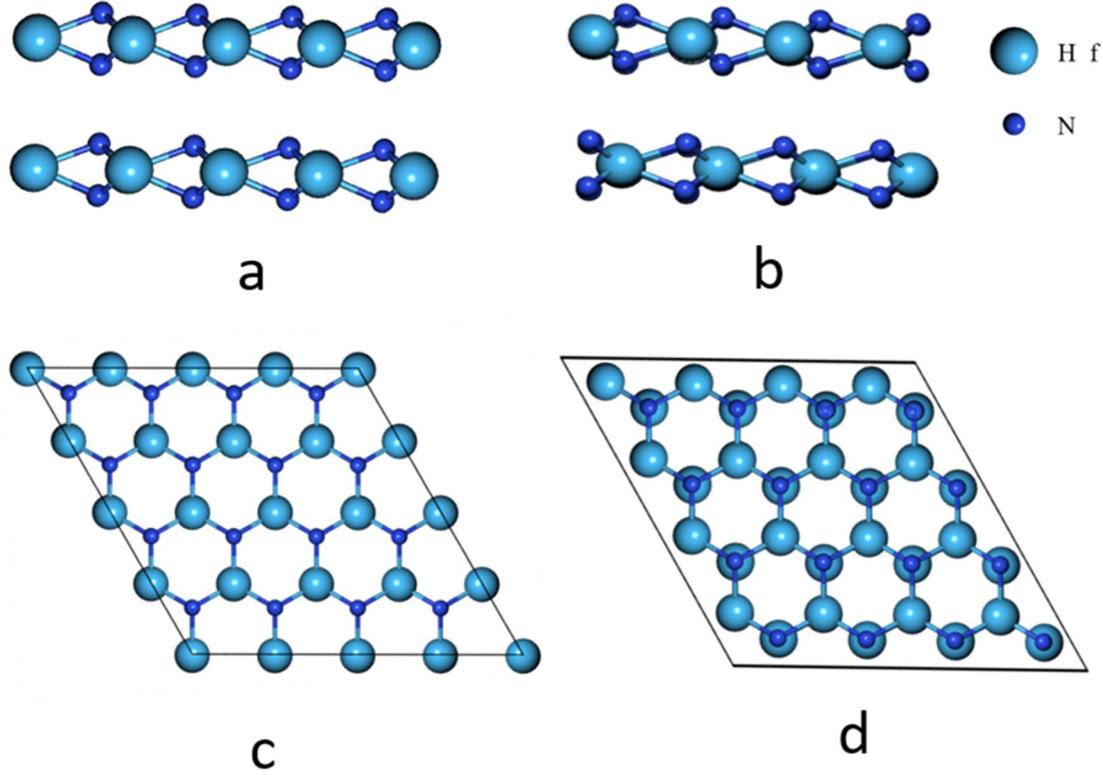

**Fig.1** Side view of (a) AA stacking, (b) AB stacking bilayer. Top view of 4×4×1 supercell of HfN$_2$ bilayer (c) AA and (d) AB stacking.

We have calculated cohesive energy to check the feasibility of two-dimensional monolayer and bilayer HfN$_2$ by using the following equation,

$$E_{ch} = \frac{(nE_{Hf} + 2nE_N) - E_t}{3n} \quad (5)$$

Where n is number of layers. $E_{Hf}$, $E_N$, $E_t$ are the energy of isolated Hf, N atoms and total energy respectively. The positive value of cohesive energy for monolayer (8.21 eV) and bilayer (8.75 eV for both stacking) suggests the stability of the material.

*Electronic properties:*

The electronic properties of monolayer and bilayer HfN$_2$ were studied from the band structure. The band structure of HfN$_2$ monolayer along high symmetric K path Γ-K-M-Γ in hexagonal Brillouin zone is shown in figure 2 (a) which shows that HfN$_2$ has a bandgap of 1.50 eV. The valance band maxima (VBM), conduction band minima (CBM) lie at K point and Fermi level rests within the gap. The electronic band structure for HfN$_2$ bilayers are shown in figure 2(b) and 2(c). VBM and CBM are at K point which means the bilayers also have direct bandgap. The bandgap for AA stacking bilayer is 1.29 eV whereas AB stacking has a gap of 1.45 eV, and is less as compared to monolayer. The reduction of band gap of AA bilayer is due to splitting of bands of two layers which is small for AB stacking. Stacking dependent band splitting is due to nearest-neighbor atomic interaction of intralayer atoms as already explained by Ho et al. for bilayer graphite[40]. The projected density of states of HfN$_2$ monolayer is shown in supplementary information (Fig.S1) gives information about the contribution in conduction and valance band. VBM is contributed by p$_y$, p$_z$ orbital of N atoms and d$_{zy}$ orbital of Hf atom whereas CBM is contributed by d$_{xy}$, $d_{z^2}$ orbital of Hf atom and p$_x$ orbital of N atoms. The spin orbit coupling (SOC) effect on band structures has also been studied and is shown in figure 2(d)-(f). Calculations have been carried out with SOC and non-collinear version is also included and bandgap of 1.31 eV, 1.05 eV and 1.26 eV are observed for monolayer, AA and AB bilayer respectively. All the bands are spin split at high symmetric K point. The conduction band's splitting is more prominent than the valance band's splitting because there is more contribution of d orbital of Hf atom which is mainly responsible for band splitting and maximum band splitting of monolayer occurs at CBM (313.3 meV) than VBM (42.2 meV).

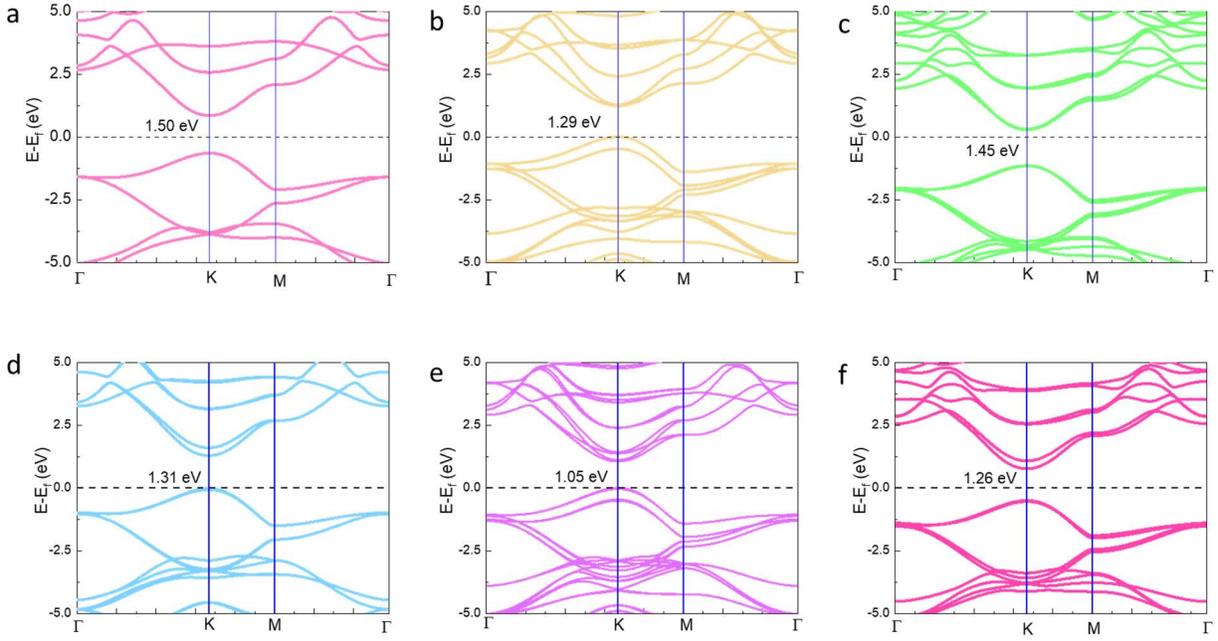

**Fig.2:** Band structure of $HfN_2$ (a) monolayer, bilayer (b) AA stacking (c) AB stacking without SOC. SOC effect on band structure of $HfN_2$ (d) monolayer (e) & (f) AA and AB bilayer $HfN_2$.

Bardeen-Shockley's deformation potential theory was used to calculate the mobility of charge carriers[41]. The deformation potential was calculated from the slope of VBM and CBM changes with applied small strain. The mobility was calculated from the following equation

$$\mu_{2D} = \frac{2e\hbar^3 C_{2D}}{3K_B T (m^*)^2 E_{dp}^2}$$

Where, $E_{dp}$ is deformation potential calculated from the energy band edge changed with applied strain (Fig. S2), $m^*$ is effective mass and $C_{2D}$ is elastic constant of 2D materials. The mobility of $HfN_2$ monolayer, AA and AB stacking bilayer are 211.3 cm$^2$/(Vs) (for electron), 254.7 cm$^2$/(Vs) (for hole), 363.2 cm$^2$/(Vs) (for electron), 522.6 cm$^2$/(Vs) (for hole) and 549 cm$^2$/(Vs) (for electron), 868 cm$^2$/(Vs), respectively.

## *Optical properties:*

Optical properties calculation of monolayer and bilayer HfN$_2$ were performed along the normal to the plane (Z-direction) using SIESTA package[35]. Figure 3(a)-3(c) and 3(d)-3(f) show the dielectric function, absorption coefficient, refractive index variation of monolayer and bilayers with photon energy. The $\varepsilon_1$ value of monolayer has lower value than that of bilayer HfN$_2$. Zero-point value of real part of monolayer is 1.72 and increases continuously to 2.20 till the first peak at 4.14 eV. A second peak of maximum value 2.64 was observed at 6.54 eV. Bilayer material also followed the similar trend as monolayer with slightly higher value. Zero point, first peak and second peak values of real part are 2.35, 3.59, 3.78 for AA and 2.35, 3.72, 3.66 for AB stacking bilayer at 0, 4.31 eV, 6.48 eV and 0, 4.50 eV, 6.18 eV respectively. A small negative value has been found between 8.21 eV to 8.55 eV energy region for monolayer and between 7.49 eV and 8.20 eV for bilayer. At high energy range the real parts are almost stable. The imaginary parts of dielectric function ($\varepsilon_2$) have negligible value till 3.75 eV for monolayer and after that it increases to the highest value of 0.58 at 5.1 eV. A second peak arises at 7.95 eV and starts decreasing to nearly zero at 20 eV. Highest values of 4.56 and 3.96 was found at 7.05 eV for AA and AB bilayer. The absorption coefficient is negligible till 3.84 eV and after that it increases sharply to a maximum value of $8.81 \times 10^5$ cm$^{-1}$ at 8.15 eV for monolayer and after this point decrease of absorption coefficient was observed which reached to a minimum value of $3.53 \times 10^5$ cm$^{-1}$ at 8.84 eV. A second peak was observed at 9.90 eV and afterward absorption decreases. The absorption peak for monolayer and bilayer HfN$_2$ has been observed in the ultraviolet region, so this material can be used as ultraviolet optoelectronic devices such as photodetector, photodiode etc. The zero-point refractive index ($\eta$) of monolayer and bilayer are 1.31 and 1.53 respectively. $\eta$ increases with the increase in energy and reaches the highest value of 1.65 (for monolayer), 2.06 and 2.01 (for AA and AB bilayers) at 6.54 eV. At higher energy range minimal change of refractive index was observed.

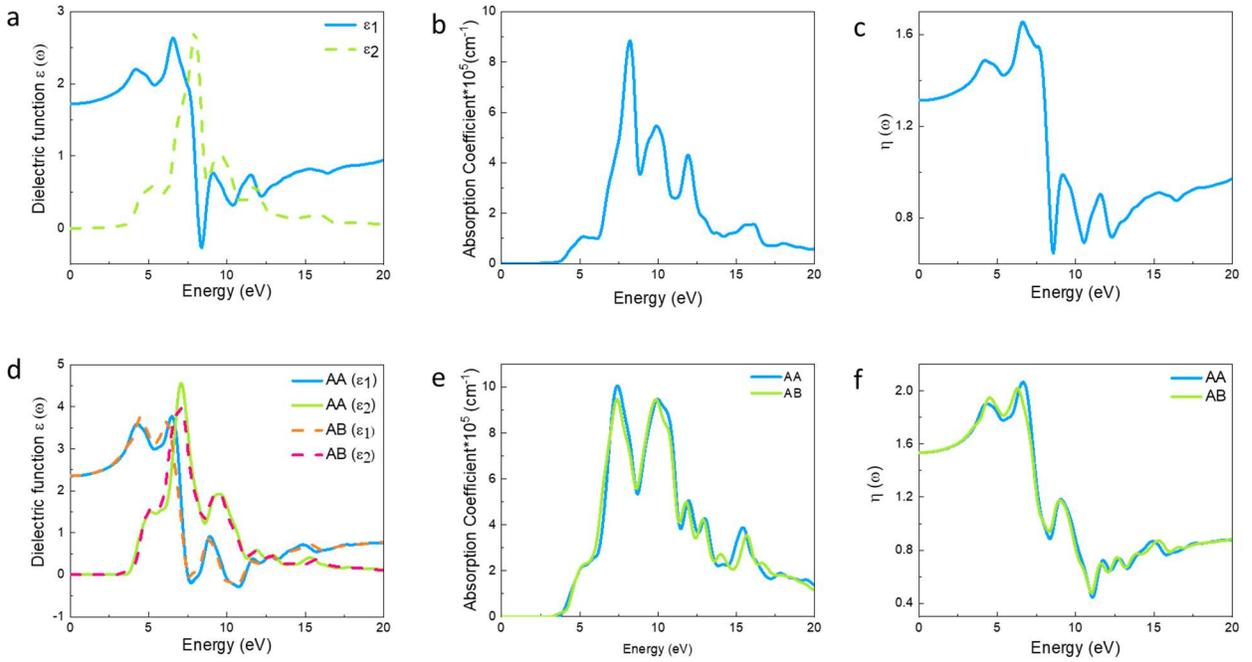

**Fig. 3:** For monolayer of HfN$_2$ (a) ε$_1$(solid line) and ε$_2$ (dashed line) (b) Absorption coefficient (c) Refractive index variation with photon energy. (d) ε$_1$ and ε$_2$(e) absorption coefficient (f) refractive index variation of AA and AB stacking bilayers of HfN$_2$.

## *Mechanical and Piezoelectric properties:*

The mechanical strength of both monolayer and bilayer HfN$_2$ was studied by calculating the elastic properties of the material. The elastic constants C$_{ij}$ (i, j=1, 2) in x and y directions can be calculated from the slope of energy vs strain plot (Fig. S3). The value of elastic constants are C$_{11}$=C$_{22}$=155.6 N/m and C$_{12}$=C$_{21}$=92.3 N/m for monolayer which is nearly same as reported value[31] and C$_{11}$=C$_{22}$=535.7 N/m and C$_{12}$=C$_{21}$=429.9 N/m for bilayer. The Born criterion of mechanical stability has been checked using the equation $C_{11}^2 - C_{12}^2$ which has positive value and thus assures the stability. The Young's moduli of the materials have been calculated from the equation $Y = (C_{11}^2 - C_{12}^2)/C_{11}$. High and positive value of elastic constants suggest strong bonding between atoms of the material. All the calculated parameters are listed in Table1.

Piezoelectric properties of monolayer and bilayer HfN$_2$ are calculated using the well-known Duerloo's theoretical approach for two dimensional materials[42]. The piezoelectric stress tensor (e$_{ij}$) and piezoelectric strain tensor (d$_{ij}$) were calculated from the following equations

$$e_{ij} = \left(\frac{\partial P_i}{\partial \varepsilon_j}\right) \quad (6)$$

$$d_{ij} = \left(\frac{\partial P_i}{\partial \sigma_j}\right) \quad (7)$$

Where, $P_i$ is the polarization induced along i direction when a strain is applied along a certain direction j (i, j=1, 2 which represent x, y direction), provided with constant or zero electric field and constant temperature. $\varepsilon_j$ is the applied strain along j direction and $\sigma_j$ is corresponding stress along the same direction. Strain tensor ($d_{11}$) and stress tensor ($e_{11}$) can be correlated with elastic stiffness constants as

$$d_{11} = \frac{e_{11}}{C_{11}-C_{12}} \quad (8)$$

Where, $C_{11}$ and $C_{12}$ are the stiffness constants. Piezoelectric coefficient $e_{11}$ can be calculated from the equation

$$P_1(\varepsilon_{11}) - P_2(\varepsilon_{11} = 0) = \varepsilon_{11} e_{11} \quad (9)$$

Figure 4 shows the polarization change with applied strain for monolayer and bilayer $HfN_2$ which is changing linearly with strain. $e_{11}$ can be calculated by least square fitting and taking the slope of the plot. For monolayer the value of $e_{11}$ is $4.63 \times 10^{-10}$ C/m whereas for bilayer the value increases to $7.96 \times 10^{-10}$ C/m (AA stacking) and $8.97 \times 10^{-10}$ C/m (AB stacking). The values of $d_{11}$ are 7.31 pm/V, 7.52 pm/V and 12.59 pm/V for monolayer and bilayers (AA and AB) respectively. So, bilayer $HfN_2$ has better piezoelectric property than monolayer.

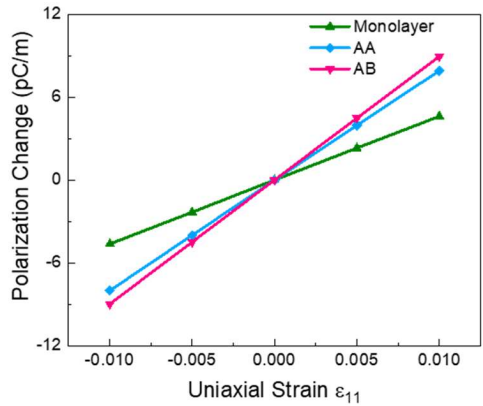

**Fig.4:** Polarization change of monolayer and bilayer $HfN_2$ with applied strain.

| Material | $e_{11}$ ($10^{-10}$ C/m) | $d_{11}$ (pm/V) | $C_{11}$ (N/m) | $C_{12}$ (N/m) | Y (N/m) | Reference |
|---|---|---|---|---|---|---|
| $MoS_2$, $MoSe_2$, $MoTe_2$ | 3.06, 2.80, 2.98 | 2.91, 3.05, 4.33 | 153, 131, 101 | 48, 39, 32 | 138, 119, 91 | Reference[42] |
| $WS_2$, $WSe_2$, $WTe_2$ | 2.20, 1.93, 1.60 | 1.93, 1.80, 1.88 | 170, 147, 116 | 56, 40, 31 | 151, 136, 107 | Reference[42] |
| CdO, MgO | 3.18, 5.61 | 23.8, 27.4 | 83, 98 | 29, 29 | 72, 89 | Reference[43] |
| GaS, GaSe, InSe | 5.39, 5.22, 5.17 | 8.29, 9.67, 13.26 | 108, 91, 75 | 32, 26, 35 | 98, 83, 59 | Reference[44] |
| $Ga_2SSe$, $In_2SSe$ | 4.09, 7.94 | 5.15, 13.06 | 112, 88 | 33, 27 | 102, 79 | Reference[45] |
| $HfN_2$ | 4.63 | 7.31 | 155.6 | 92.3 | 100.8 | This Work |
| $HfN_2$ bilayer AA/AB stacking | 7.96/8.97 | 7.52/12.59 | 535.7/493.8 | 429.9/422.6 | 190.7, 132 | This Work |

**Table 1:** Comparisons between piezoelectric coefficients and elastic constants of some popular monolayer materials as reported previously and our calculated value for $HfN_2$ monolayer and bilayer.

## *Thermoelectric properties:*

Thermoelectric properties of monolayer and bilayer $HfN_2$ were calculated using BoltzTraP code[38]. The Seebeck coefficients of monolayer, AA stacked, and AB stacked bilayers at 300K, 600K and 900K of $HfN_2$ vary with chemical potential and are shown in figure 5(a)-5(c). At 300K Seebeck coefficient has its maximum value as compared to 600K and 900K as it is inversely proportional to the temperature gradient. At 300K highest observed values are 2353 $\mu VK^{-1}$, 1945 $\mu VK^{-1}$ and 2302 $\mu VK^{-1}$ for p-type ($\mu<0$) monolayer, bilayer AA and AB respectively whereas for n-type ($\mu>0$) these values are -2426 $\mu VK^{-1}$, -2010 $\mu VK^{-1}$ and -2267 $\mu VK^{-1}$ respectively. The higher value of Seebeck coefficient of monolayer than AA stacking bilayer can be easily explained from

Goldsmid and Sharp's simple estimation that is proportional to band gap of the material[46]. The change of electrical conductivity per relaxation time with chemical potential for monolayer and bilayers are shown in fig 5(d)-5(f) which clearly shows that electrical conductivity has negligible dependency on temperature. At band edge, the electrical conductivity of n-type carriers is more than that of p-type carriers. The highest values of electrical conductivity of n-type carriers at band edge for the materials are $0.67 \times 10^{19}$ S/m, $2.55 \times 10^{19}$ S/m and $1.68 \times 10^{19}$ S/m respectively. Relaxation time scaled power factor (PF) varies with chemical potential and is shown in fig 5(g)-5(i). At 300K, PF of monolayer is $6.82 \times 10^{10}$ Wm$^{-1}$K$^2$s$^{-1}$ for n-type and $4.16 \times 10^{10}$ Wm$^{-1}$K$^2$s$^{-1}$ for p-type carriers. Highest value of $15.47 \times 10^{10}$ Wm$^{-1}$K$^2$s$^{-1}$ for n-type materials and $12.84 \times 10^{10}$ Wm$^{-1}$K$^2$s$^{-1}$ for p-type carrier has been observed at 900K. For all three cases as there are broad peaks along negative chemical potential side i.e. near VBM, so first peak which is near the band edge was taken into account. Bilayers have greater electrical conductivity, so it is obvious that PF must be greater than monolayer. Highest value for n-type charge carriers at 900K are $20.62 \times 10^{10}$ Wm$^{-1}$K$^2$s$^{-1}$ and $24.13 \times 10^{10}$ Wm$^{-1}$K$^2$s$^{-1}$ near band edge whereas for p-type carriers these values are $10.14 \times 10^{10}$ Wm$^{-1}$K$^2$s$^{-1}$ and $21.75 \times 10^{10}$ Wm$^{-1}$K$^2$s$^{-1}$ respectively. The optimized thermoelectric properties for monolayer and bilayer revealed that these can be used to fabricate good quality thermoelectric devices.

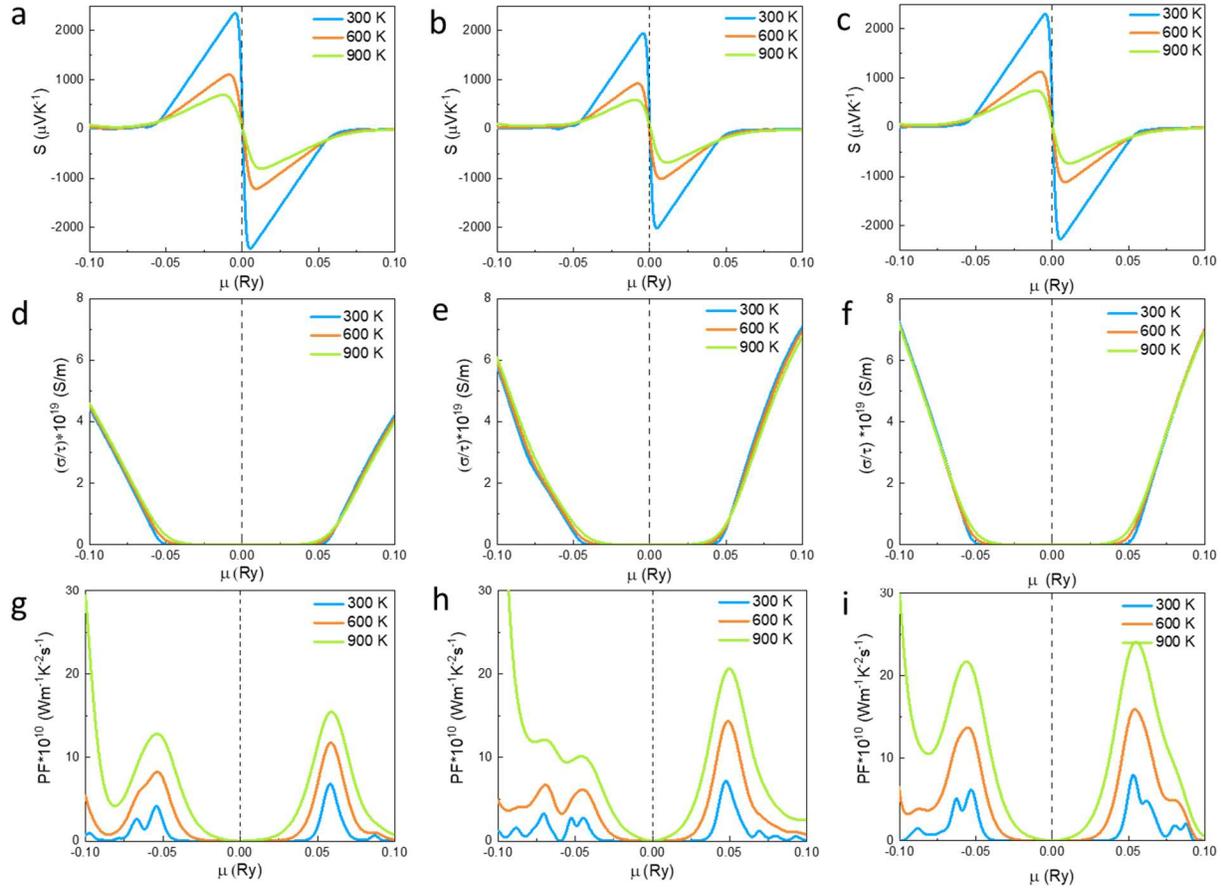

**Fig. 5:** (a)-(c) Variation of S (d)-(f) $\sigma/\tau$ (g)-(i) $S^2\sigma/\tau$ with chemical potential for monolayer & bilayer with AA and AB stacking $HfN_2$.

## *Thermal conductivity:*

Thermal conductivity plays a crucial role on thermoelectric properties of material as it is inversely proportional to the thermoelectric efficiency, popularly known as figure of merit. So, to get better efficiency thermal conductivity should have lower value. Contribution to the thermal conductivity is due to the electron and phonon or lattice vibration. Figure 6(a) indicates the change of lattice thermal conductivity ($K_{ph}$) of monolayer, and bilayers of different stacking with temperature. Although at lower temperature $K_{ph}$ increases and reached a maximum and then decreases at higher temperatures but we are interested about higher temperatures only. The $K_{ph}$ of bilayers is slightly smaller than that of monolayer. At room temperature (300K) values of $K_{ph}$ are 4.875 W/(mK), 3.641 W/(mK) and 2.63 W/(mK) for monolayer and bilayers (AA and AB stacked) respectively. Reduction of thermal conductivity with increasing number of layers has been reported in many previous works[47,48]. The reason of smaller value can be understood by looking at mean free path and lifetime of phonon as shown in figure 6(b) and 6(c). Decreasing behavior of thermal conductivity can be explained by modified Slack model[49] for non-metallic material which suggests that it is inversely proportional to absolute temperature.

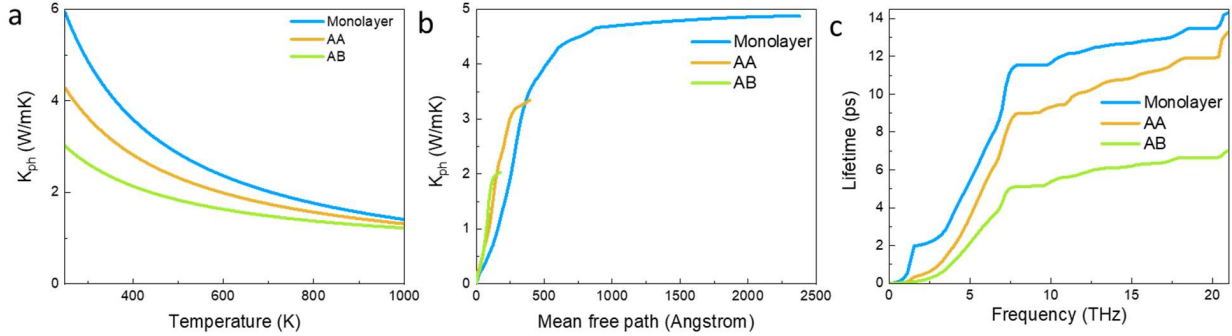

**Fig. 6:** (a) $K_{ph}$ variation with temperature (b) $K_{ph}$ with variation of mean free path (c) phonon lifetime varies with phonon frequency for monolayer & bilayer $HfN_2$.

Phonon thermal conductivity varies with mean free path (MFP) of phonon at room temperature and has been shown for all three types of materials. The MFP expression in terms of group velocity and relaxation time is given by

$$\Lambda_\lambda = V_\lambda \times \tau_\lambda$$

Where $V_\lambda$ is the group velocity of phonon mode $\lambda$ and $\tau_\lambda$ is the corresponding lifetime of phonon. The maximum values of MFP at 300K are 2378Å for monolayer, 394Å and 178Å for bilayers. The decrease in MFP with the increase in temperature is due to the increase in scattering probability that leads to lower MFP. Lifetime of phonon varies with frequency of lattice vibration shown in figure for monolayer and bilayers indicates that monolayer has greater phonon lifetime than that of bilayers. At room temperature highest lifetimes of 14.42ps, 13.56ps and 7.34ps have been observed for monolayer and AA, AB bilayers respectively and they decrease with temperature. Due to lower phonon lifetime and MFP value of bilayer $HfN_2$, the $K_{ph}$ value of bilayers has comparatively lower value than that of monolayer $HfN_2$. Another contribution to the thermal conductivity is due to the electrons and is shown in Fig. S4. Relaxation time scaled thermal conductivity varies with chemical potential at different temperatures and is plotted in the figure.

## Figure of merit:

Efficiency of thermoelectric materials can be defined in terms of thermoelectric figure of merit or popularly called as ZT of material. ZT can be expressed by the equation

$$ZT = \frac{S^2 \sigma T}{K_e + K_{ph}}$$

Where $K_e$, $K_{ph}$, S, σ, T are the electronic thermal conductivity, lattice thermal conductivity, seebeck coefficient, electrical conductivity and absolute temperature, respectively. ZT values have been calculated using this equation from the calculated parameters and is shown in figure 7. Higher value of ZT makes $HfN_2$ a competitor to the well-established thermoelectric materials. Monolayer has showed the highest ZT of 0.15 (0.10), 0.49 (0.37) and 0.65 (0.56) for n-type (p-type) material at 300K, 600K and 900K respectively. Bilayer $HfN_2$ has shown even better results than that of monolayer as power factor of bilayers has higher value than that of monolayer. The two factors that contributed to the high ZT are comparatively lower electronic thermal conductivity ($K_e$) and lattice thermal conductivity ($K_{ph}$). Higher value in the positive chemical potential region suggests that the n-type doping will be more effective than that of p-type doping to enhance thermoelectric properties. Comparison of figure of merit of some transition metal dichalcogenides (TMDC) with our calculated values is listed in table 2.

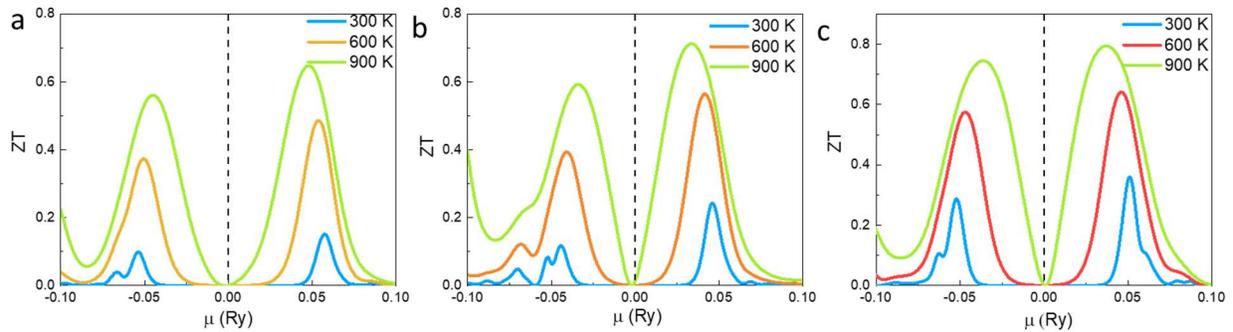

**Fig.7:** Thermoelectric figure of merit of (a) $HfN_2$ monolayer, $HfN_2$ bilayers with (b) AA stacking (c) AB stacking

| Material | ZT | | Temperature |
|---|---|---|---|
| | n-tpe | p-type | |
| $WS_2$ monolayer | 0.90 | 0.77 | 1500K *[a] |
| $WSe_2$ monolayer | 0.75 | 0.5 | 500K *[b] |
| $MoS_2$ monolayer | 0.26 | 0.16 | 500K *[c] |
| (Zr/Hf)$Se_2$ monolayer | 0.95 | 0.87 | 600 K *[d] |
| $HfN_2$ monolayer | 0.65 | 0.56 | 900K [This work] |
| $HfN_2$ bilayer AA/AB | 0.71/0.80 | 0.59/0.75 | 900K [This work] |

[a] reference[20]. [b] reference[50]. [c] reference[50]. [d] reference[51]. (*) Represents highest ZT value with finite doping concentration.

**Table 2:** Comparison of figure of merit between different TMDCs and TMDNHfN$_2$.

## *Conclusions:*

We have calculated the optoelectronic, piezoelectric and thermoelectric properties of monolayer and bilayer $HfN_2$ with AA and AB stacking using density functional theory. Both monolayer and bilayer are direct band gap semiconductors having VBM and CBM at high symmetric K point. Higher values of absorption coefficient in ultraviolet region indicates that $HfN_2$ can be used as optoelectronic device like ultraviolet photodetector. Piezoelectric properties were studied using Duerloo's theoretical approach for two dimensional materials. Values of piezoelectric stress tensor and strain tensor are $4.63 \times 10^{-10}$ C/m and 7.31 pm/V for monolayer whereas for bilayers those values are $7.96 \times 10^{-10}$ C/m, $8.97 \times 10^{-10}$ C/m and 7.52 pm/V, 12.59 pm/V. Monolayer has lower power factor than that of bilayers and because of that bilayers show greater thermoelectric efficiency.

## *Acknowledgement:*

The authors are thankful to the Ministry of Human Resource and Development (MHRD) for the funding and Indian Institute of Technology Jodhpur for the computational infrastructure.

The data that support the findings of this study are available from the corresponding author upon reasonable request.